\documentclass{vgtc}         

\ifpdf%
  \pdfoutput=1\relax
  \DeclareGraphicsExtensions{.pdf,.png,.jpg,.jpeg}
\else%
  \DeclareGraphicsExtensions{.eps}
\fi%

\usepackage[expansion=false]{microtype}
\usepackage{booktabs}
\usepackage{siunitx}
\usepackage{amsmath}

\usepackage{times}
\usepackage{mathptmx}

\vgtcpapertype{short paper}

\onlineid{0}

\vgtcinsertpkg

\usepackage{xurl}
\usepackage{orcidlink}

\hypersetup{
  pdftitle={Information Terra: A Narrative-Anchored Semantic-First Projection of Document Embeddings},
  pdfauthor={Brian Keith-Norambuena, Fausto German, Chris North},
  pdfkeywords={Narrative visualization, information landscapes, document embeddings, spherical projection},
  pdflang={en}
}

\begin{document}

\title{Information Terra: A Narrative-Anchored Semantic-First
       Projection of Document Embeddings}

\author{%
  Brian Keith-Norambuena\,\orcidlink{0000-0001-5734-8962}\thanks{e-mail: brian.keith@ucn.cl}\\%
  \scriptsize Universidad Cat\'olica del Norte, Chile%
  \and Fausto German\,\orcidlink{0009-0005-0954-4578}\thanks{e-mail: fgermanj@vt.edu}\\%
  \scriptsize Virginia Tech%
  \and Chris North\,\orcidlink{0000-0002-8786-7103}\thanks{e-mail: north@cs.vt.edu}\\%
  \scriptsize Virginia Tech%
}


\shortauthortitle{Keith-Norambuena \MakeLowercase{\textit{et al.}}: Information Terra}

\abstract{%
We introduce \emph{Information Terra}, a narrative-anchored
semantic-first projection that places a document corpus on an
Earth-like globe whose poles are two user-chosen endpoint
documents and whose prime meridian is the great-circle geodesic
between them on the embedding hypersphere---so latitude encodes narrative progress and longitude thematic deviation.  Land features are recovered from document density via kernel density estimation and labeled by theme. A \textit{narrative trail} built from the underlying narrative coherence graph, and constrained to be monotone in geodesic progress, provides a readable storyline. The projection's axes are semantically grounded in the user's chosen narrative endpoints, and the globe metaphor affords rotation and antipodal reading. We demonstrate the method on a 540-article Cuban Protests corpus, showing a storyline from Obama's 2016 visit to the 2021 International Aid during the protests.%
}


\CCScatlist{
 \CCScatTwelve{Human-centered computing}{Visualization}{Visualization techniques}{Information visualization};
 \CCScatTwelve{Human-centered computing}{Visualization}{Visualization design and evaluation methods}{}%
}

\teaser{%
  \centering
  \includegraphics[alt={Three views of the document globe for the Cuba--U.S. corpus. Colored landmasses cover the sphere, and a black narrative path runs from the source event at the south pole to the target event at the north pole along a red dashed prime meridian, shown tilted, from the side, and from the back.},width=\linewidth]{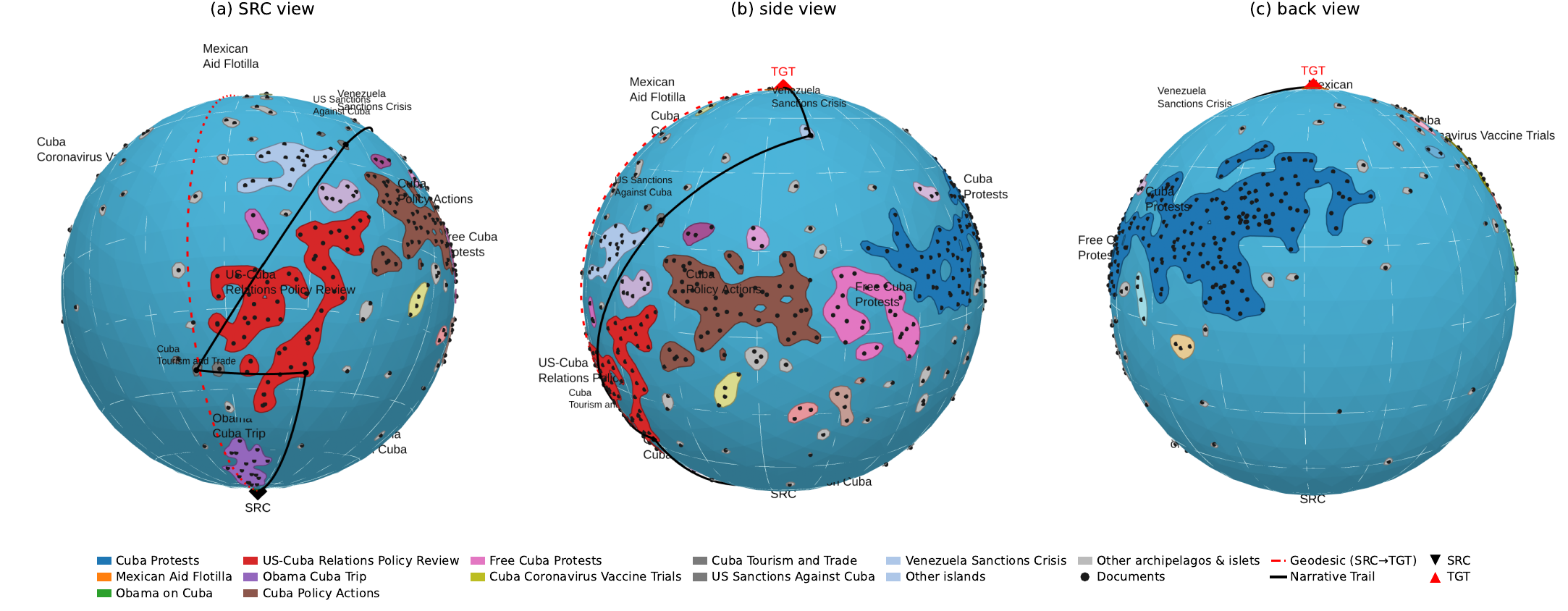}
  \caption{\textbf{Information Terra.} Reference narrative from Obama's comments on Cuba (source event, south pole) to the 2021 Mexican aid flotilla (target event, north pole) on a 540-article Cuba-U.S. corpus, rendered as a globe seen from three angles: \textbf{(a)} tilted to show the storyline from its start, \textbf{(b)} a side view surfacing continents away from the prime meridian, \textbf{(c)} a back view with the storyline on the far hemisphere. Landmasses are named by a small local language model; the black curve is the geodesic-monotone narrative path, and the red dashed meridian is the SLERP geodesic between the endpoints. Smaller landmasses are grayed, with a darker shade for those the narrative visits. A demo for the system is available at: \url{https://huggingface.co/spaces/briankeithn/InformationTerra}}
    \label{fig:teaser}
}

\maketitle

\section{Introduction}\label{sec:intro}
Narrative visualization and extraction span a broad methodological space~\cite{keith2023survey}. We work in the \emph{event-based narrative extraction} paradigm, in which each document is treated as a single event and a narrative is a temporally ordered, thematically coherent sequence of such document-events. Methods in this paradigm extract narratives as discrete structures---chains, directed acyclic graphs, and maximum capacity paths---on a coherence graph; recent extensions steer the substrate via user agendas~\cite{keith2026agenda} or semantic interaction~\cite{keith2026semantic}.

In several of these methods a projection of the embedding space is used in the computation of the coherence function itself, so the projection's axes shape which paths are found, not merely how they are drawn. Rendering is a separate choice: extracted structures are most often shown as abstract graph diagrams or timelines, and the underlying projection can also be displayed directly, though its axes carry no narrative meaning.  Information Terra addresses that substrate: a \emph{semantic-first} projection whose axes are fixed by the analyst's chosen narrative endpoints before extraction or rendering, serving both as the coherence substrate and as the layout on which the result is read, as shown in Figure \ref{fig:teaser}.

\begin{figure}[tb]
\centering
\includegraphics[alt={Two flat scatterplots of the same corpus, a UMAP layout on the left and a PCA layout on the right. In both, a red dashed geodesic and a black path loop outside the dense cloud of document points.},width=\columnwidth]{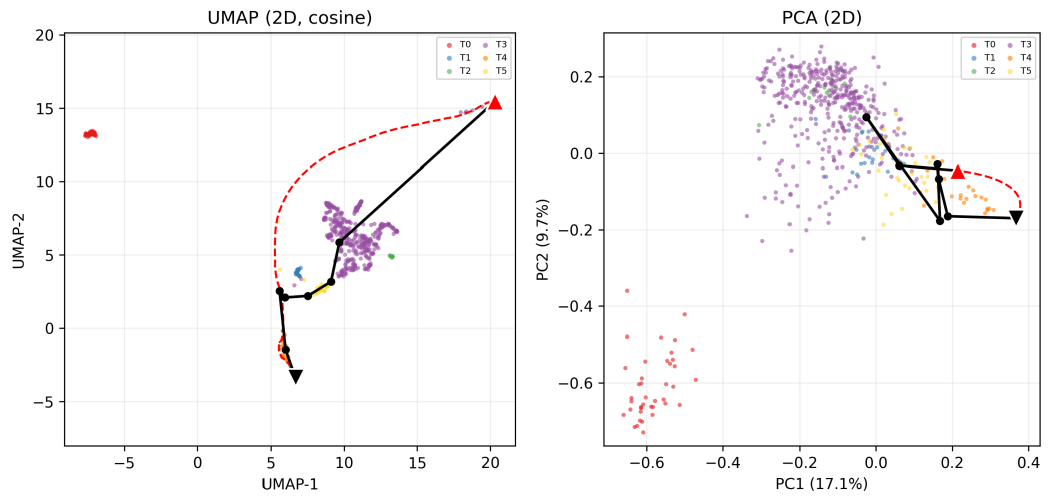}
\caption{\textbf{Why the globe helps.} Flat UMAP~\cite{umap} (left) and PCA (right) of the same Cuba-U.S. corpus, with the SLERP geodesic (red dashed) and unconstrained MCP (black) overlaid. The axes are chosen by topology (UMAP's neighborhood graph) or variance (PCA's principal components), not by the intended narrative structure; as a result the geodesic loops outside the document cloud in both cases. These layouts come from variance- and topology-driven axes. The Mollweide unfolding of the Information Terra projection gives the endpoint-anchored flat view used elsewhere in the paper.}
\label{fig:compare}
\end{figure}

Prior methods use flat projections (UMAP, PCA) whose variance- or neighborhood-driven axes are chosen independently of endpoint pairs, so the spherical linear interpolation (SLERP) geodesic between a user's starting event (\textit{SRC}) and ending event (\textit{TGT}) is visible only as an overlay (Figure \ref{fig:compare}).

We argue that for narrative tasks, the \emph{embedding sphere itself} provides the right coordinate system once two endpoints are fixed.  Normalized embeddings live on \(S^{d-1}\); the SLERP great-circle arc between two endpoint documents is a geodesic on that sphere. We take that geodesic as the prime meridian of a bipolar coordinate system, with SRC at the south pole and TGT at the north pole, and project every document to a \((\mathrm{lat}, \mathrm{lon})\) pair in which latitude encodes geodesic progress and longitude encodes thematic deviation from the shortest path (the geodesic itself). We call the resulting system \emph{Information Terra}. The system is built in two stages. A \emph{projection} stage places the documents on the sphere and extracts a non-backtracking narrative path. A \emph{visualization} stage turns the resulting sphere into an Earth-like map with continents, coastlines, and labels.

Our contributions, in order of the pipeline, are as follows: \textit{(i)}  A deterministic, closed-form geodesic-anchored bipolar projection from \(S^{d-1}\) onto a 2-sphere based on predefined narrative endpoints. \textit{(ii)} A variant of Narrative Trails that modifies the Maximum Capacity Path (MCP) algorithm to produce monotone paths along the geodesic axis, ensuring no narrative backtracking on the projection. \textit{(iii)} A KDE-based landmass-extraction and rendering pipeline of the projection that generates a labeled Earth-like map based on the geodesic-anchored bipolar projection.

\section{Related Work}\label{sec:related}
\emph{Information landscape visualizations.} Several systems render document corpora as geographic landscapes for structure discovery: WEBSOM~\cite{websom} and IN-SPIRE ThemeView~\cite{inspire_themeview} organize documents by learned topology or force-directed proximity; Topic Islands~\cite{topic_islands} and ThemeRiver~\cite{themeriver,themeriver_tvcg} extend the idiom to terrain and temporal flow; WebLyzard~\cite{weblyzard_infolandscape} applies it incrementally to web-scale media monitoring. In all cases layout axes are determined by document statistics---variance, topology, or time---with no mechanism to align the layout with a chosen endpoint pair. Information Terra shares the landscape idiom but fixes its axes by narrative endpoints, so latitude encodes geodesic progress and longitude encodes thematic deviation, giving every map region an explicit semantic interpretation rather than an emergent one.

\emph{Narrative extraction.} Within the event-based paradigm~\cite{keith2023survey}, chain extractors~\cite{shahaf_connectingdots,hossain_connectingdots}, DAG-based approaches such as Metro Maps~\cite{shahaf_metromaps}, Narrative Maps~\cite{narrativemaps_csci}, and Narrative Trails~\cite{narrativetrails}---a Maximum Capacity Path formulation we build on directly---extract narratives as discrete structures; storyline layouts~\cite{storyflow} and narrative-map design principles~\cite{narrativemaps_design} address rendering and sensemaking quality. Recent extensions steer the substrate via user agendas~\cite{keith2026agenda} or semantic interaction~\cite{keith2026semantic}; the Interactive Narrative Analytics framework~\cite{keith_ina} argues that tight coupling between extraction and visualization is essential for sensemaking. Information Terra operationalizes this coupling geometrically: the narrative-anchored projection simultaneously defines the coherence substrate and provides the rendering coordinates in terms of narrative progress.

\emph{User-driven projections.} Semantic interaction for visual text analytics~\cite{endert_si_chi2012} reshapes a document spatialization iteratively by interpreting user drags; Information Terra takes endpoints as fixed inputs and produces a deterministic geometric projection. Spherical projections on \(S^{d-1}\), studied in directional statistics~\cite{fisher_spherical_1987, mardia_directional_2000}, typically pick poles by variance, density, or convention rather than by narrative endpoints. SLERP~\cite{slerp_shoemake} supplies the great-circle primitive we repurpose as the prime meridian.

\section{Projection}\label{sec:projection}
The projection stage takes the normalized embeddings and two endpoint documents and produces a bipolar \((\mathrm{lat}, \mathrm{lon})\) coordinate for every document and a narrative path from SRC to TGT that never backtracks along the geodesic. Both depend only on the embedding geometry and the endpoints, not on rendering.

\begin{figure*}[!tb]
\centering
\includegraphics[alt={Three-stage construction shown as flat Mollweide maps. Panel (a) is a scatter of documents colored by topic with a red dashed geodesic, panel (b) is a smooth density field from kernel density estimation, and panel (c) thresholds that density into colored land regions on a blue ocean.},width=\textwidth]{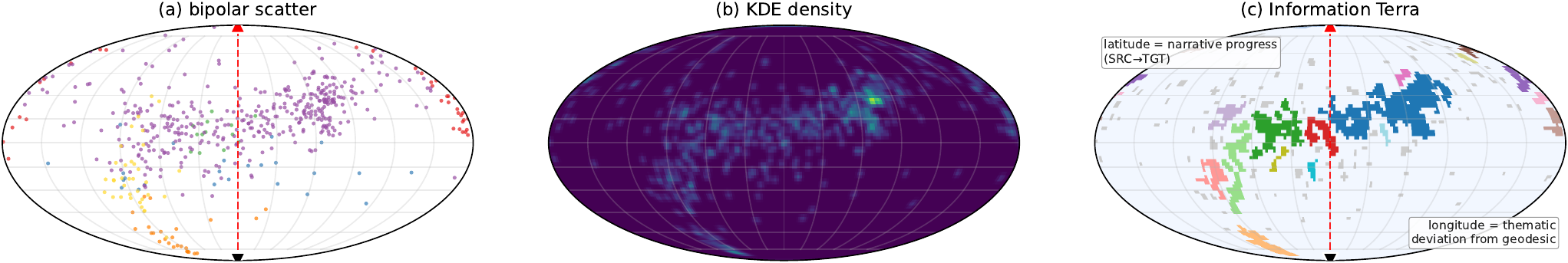}
\caption{\textbf{Pipeline of the Information Terra construction} (Mollweide unfoldings of the reference sphere).\textbf{Projection stage:} \textbf{(a)} documents in bipolar \((\mathrm{lat}, \mathrm{lon})\) coordinates, colored by HDBSCAN topic; SRC/TGT are the south/north poles and the red dashed line is the SLERP geodesic at \(\mathrm{lon}=0\). \textbf{Visualization stage:} \textbf{(b)}~Gaussian KDE on the sphere with bandwidth \(\sigma = 0.7 \times\) median nearest-neighbor distance; color encodes density. \textbf{(c)}~Land/ocean thresholding at the 65\textsuperscript{th} density percentile, with each connected component drawn in a distinct color and small features below the island threshold collapsed to gray. Latitude is narrative progress, longitude is thematic deviation from the geodesic.}
\label{fig:pipeline}
\end{figure*}

\subsection{Bipolar projection on the embedding sphere}\label{sec:bipolar}
Let \(\mathbf{a}, \mathbf{b} \in S^{d-1}\) be the two endpoint documents. The SLERP geodesic~\cite{slerp_shoemake} between them is
\begin{equation}
\gamma(t) =
  \frac{\sin\big((1-t)\Omega\big)}{\sin \Omega}\,\mathbf{a}
  + \frac{\sin(t\Omega)}{\sin \Omega}\,\mathbf{b},
\quad \Omega = \arccos(\mathbf{a}\cdot\mathbf{b}).
\end{equation}

Let \(\mathbf{x} \in S^{d-1}\) be any document embedding; its latitude is
the SLERP parameter \(t\) at which \(\gamma(t)\) is closest to \(\mathbf{x}\)
in angular distance, i.e., the \(t\) that maximizes \(\mathbf{x}\cdot\gamma(t)\).
Let \(\alpha = \mathbf{x}\cdot\mathbf{a}\) and \(\beta = \mathbf{x}\cdot\mathbf{b}\).
This maximum has closed form:
\begin{align}
t^*(\mathbf{x})  &= \mathrm{clip}_{[0,1]}\!\left(\tfrac{1}{\Omega}\,\mathrm{atan2}(\beta - \alpha\cos\Omega,\; \alpha\sin\Omega)\right),\\
\text{lat}(\mathbf{x}) &= \pi\big(t^*(\mathbf{x}) - \tfrac{1}{2}\big),
\end{align}
so latitude ranges from \(-\pi/2\) at SRC to \(+\pi/2\) at TGT, with documents past either pole mapped to the nearest pole. Scaling by \(\pi\) rather than \(\Omega\) normalizes the geodesic to span the full hemisphere for any endpoint pair, keeping the latitude axis a pure measure of narrative progress rather than a function of endpoint similarity.

For longitude, let
\(\mathbf{b}_{\perp} = (\mathbf{b} - (\mathbf{a}\cdot\mathbf{b})\mathbf{a})
                       / \|\mathbf{b} - (\mathbf{a}\cdot\mathbf{b})\mathbf{a}\|\)
be the component of \(\mathbf{b}\) orthogonal to \(\mathbf{a}\); then
\(P = [\mathbf{a} \;\; \mathbf{b}_{\perp}]\) is an orthonormal basis of
\(\mathrm{span}(\mathbf{a}, \mathbf{b})\), and
\(\mathbf{r}_{\mathbf{x}} = \mathbf{x} - P P^{\top}\mathbf{x}\) is the
residual of \(\mathbf{x}\) off that plane. Let
\(\mathbf{p}_1, \mathbf{p}_2\) be the first two principal components
of \(\{\mathbf{r}_{\mathbf{x}}\}_{\mathbf{x}}\).  Then
\begin{equation}
\text{lon}(\mathbf{x}) =
  \mathrm{atan2}\!\bigl(\mathbf{p}_2^{\top}\mathbf{r}_{\mathbf{x}},\;
                         \mathbf{p}_1^{\top}\mathbf{r}_{\mathbf{x}}\bigr),
\end{equation}
so documents lying close to the geodesic sit near \(\mathrm{lon}=0\) and thematic extremes spread along the azimuth. Longitude therefore gives the azimuthal direction in which a document deviates from the narrative geodesic, measured around the SRC--TGT axis. The residual principal axes \(\mathbf{p}_1\) and \(\mathbf{p}_2\) define the reference frame for this azimuth. Because that frame follows from the data, longitude is read by comparing documents to one another. Two documents at the same longitude share a thematic deviation direction, and two documents roughly \(180^\circ\) apart lie on opposite sides of the narrative axis. The projection is deterministic and costs \(O(N d)\) to compute \(t^*\), plus \(O(N d k)\) for the rank-\(k{=}2\) residual PCA.

\subsection{Geodesic-monotone Maximum Capacity Path}\label{sec:monotone}
Narrative Trails~\cite{narrativetrails} extracts a storyline from SRC to TGT by computing the MCP on a coherence graph---the walk that maximizes the minimum coherence edge weight. Unconstrained, the MCP can backtrack in geodesic progress (\(t^*\) decreases at a step), producing a story that visibly folds back on the globe.  We add a \emph{geodesic monotonicity} constraint: an edge \(u \rightarrow v\) is allowed only if \(t^*(v) \geq t^*(u)\) (equal-progress steps are permitted as lateral thematic moves).

\begin{figure}[t]
\centering
\includegraphics[alt={Histogram of the relative coherence loss from the geodesic-monotone constraint over 1000 random endpoint pairs. The distribution is concentrated near zero, with most pairs paying little or no cost and only a short tail beyond five percent.},width=0.75\columnwidth]{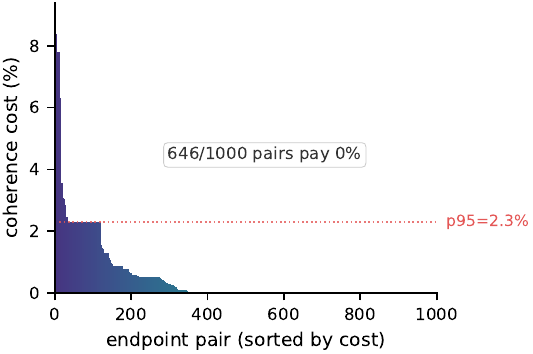}
\caption{\textbf{Distribution of geodesic-monotone MCP coherence cost} across 1000 random (SRC, TGT) pairs. Nearly 2/3 pay no cost; the tail above $2\%$ contains 118 pairs, the 95\textsuperscript{th} percentile is $2.3\%$.}
\label{fig:ablation-hist}
\end{figure}

We sampled 1000 random (SRC, TGT) pairs from the corpus and ran the MCP under both regimes. Without the monotonicity constraint, 814/1000 paths had at least one backtrack in \(t^*\). Under the constraint, the effect on coherence is practically null (Figure ~\ref{fig:ablation-hist}): 646/1000 random pairs pay no cost at all. The mean relative coherence loss is $0.5\%$, the 95\textsuperscript{th} percentile is $2.3\%$, and only 13/1000 pairs exceed $5\%$. Agenda-based Narrative Trails~\cite{keith2026agenda} reports a parallel trade-off, exchanging a small coherence drop for personalization where we exchange it for readability.

Our reference pair also sits at the p95 bottleneck edge: a $2.3\%$ cost buys a strictly forward 6-document path where the unconstrained MCP uses 8. The constrained path's absolute minimum coherence is \(0.698\), above the 82\textsuperscript{nd} percentile of our endpoint sample (median \(0.59\), p95 \(0.78\)).

\begin{figure}[t]
\centering
\includegraphics[alt={Two Mollweide maps comparing narrative paths for one endpoint pair. In panel (a) the unconstrained path dips backward in latitude several times, while in panel (b) the geodesic-monotone path rises steadily from source to target.},width=\columnwidth]{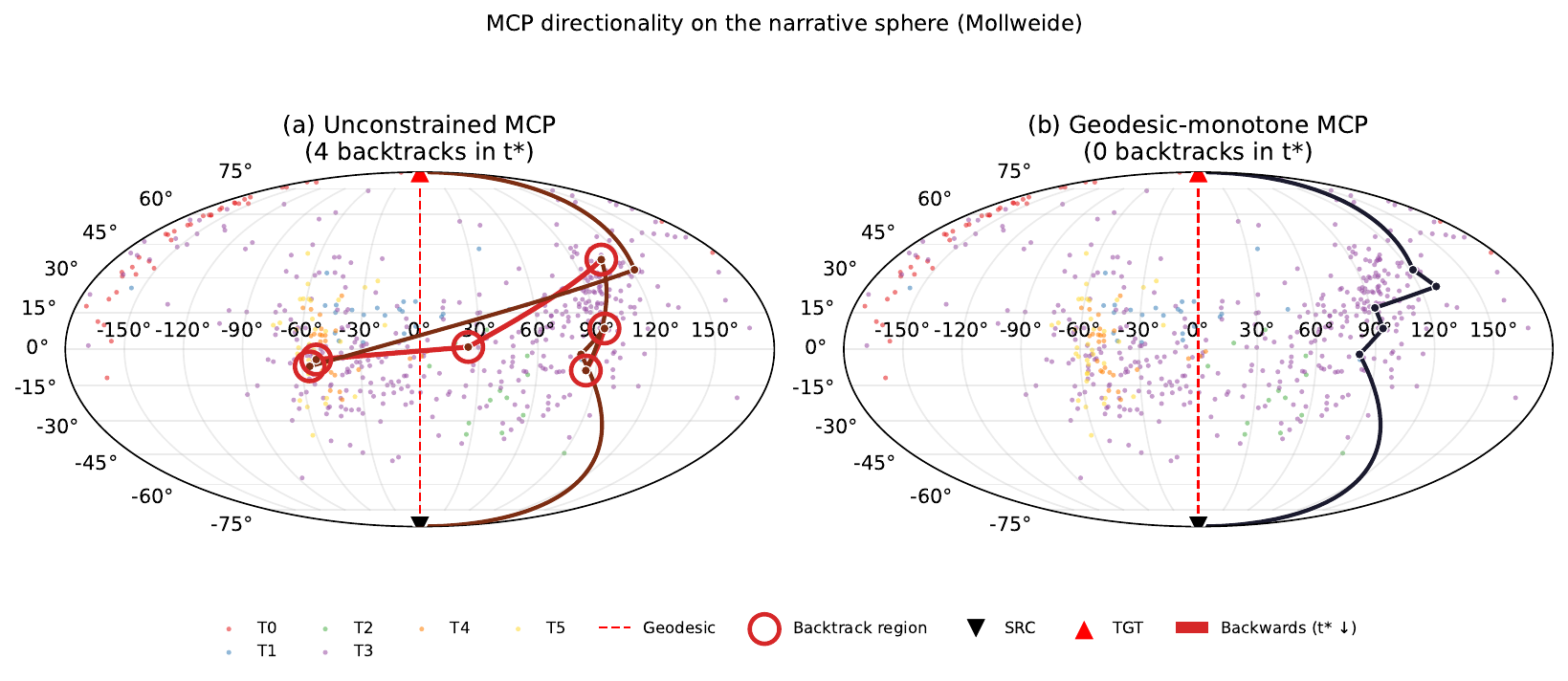}
\caption{\textbf{Geodesic directionality constraint.} Mollweide unfolding of the narrative sphere with a representative example (endpoints 166 and 338). \textbf{(a)} the unconstrained MCP backtracks in latitude four times (red segments; circled nodes mark the violated regions); \textbf{(b)} the geodesic-monotone variant advances the narrative at every step, at a $0.5\%$ relative cost in minimum edge coherence (\(0.863 \rightarrow 0.859\)).}
\label{fig:mcp-compare}
\end{figure}

\section{Visualization}\label{sec:visualization}
\begin{figure}[t]
\centering
\includegraphics[alt={A bare projected sphere showing the reference corpus as a point cloud in latitude and longitude coordinates, with the source and target documents at the two poles and no land rendering.},width=0.45\columnwidth]{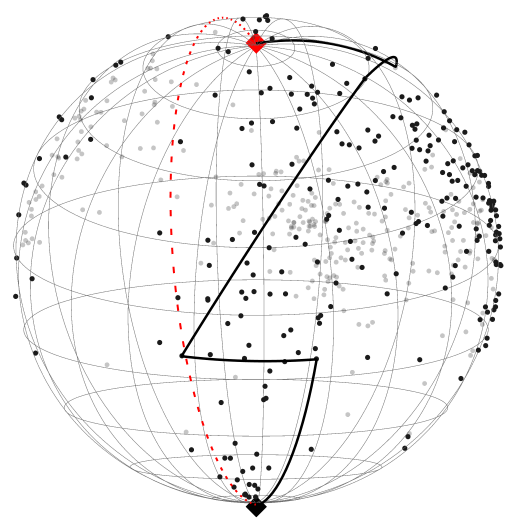}
\caption{\textbf{Projection separable from visualization.} Same \((\mathrm{lat}, \mathrm{lon})\) output as Figure ~\ref{fig:teaser} rendered without the continent overlay: documents as dots, SLERP meridian dashed red, MCP in black.}
\label{fig:bare-sphere}
\end{figure}

The proposed projection places every document at a \((\mathrm{lat}, \mathrm{lon})\) on the sphere and traces a non-backtracking narrative path from SRC to TGT. At this stage, the ``map'' is still a scatter plot (Figure ~\ref{fig:pipeline}a). The visualization stage turns it into a landscape by converting document densities into land and ocean, named with thematic labels, and rendered as the Earth-like globe or as a Mollweide unfolding (Figures~\ref{fig:pipeline} and \ref{fig:mcp-compare}). None of the decisions in this section change the projected coordinates; the rendering is separable from the method, and the same scatter supports a plain Mollweide or a 3-D point cloud if a user would prefer another visualization (Figure ~\ref{fig:bare-sphere}).

\subsection{From scatter to continents}\label{sec:continents}
To build the Earth-like rendering we discretize the sphere into a \(90 \times 180\) lat/lon grid and compute a Gaussian KDE of document density at each grid cell with haversine distances. The bandwidth \(\sigma\) is set to \(0.7\) times the median nearest-neighbor great-circle distance between documents, floored at $0.02$ rad.  Cells whose density exceeds the 65\textsuperscript{th} percentile of nonzero density are classified as land; the rest become ocean.  We also promote each document's own grid cell to land, so every document sits on land and no storyline step falls into the ocean.

Connected components are classified by cell count into \emph{continents} (\(\geq 50\) cells) and \emph{islands} (\(\geq 8\)). Smaller components form \emph{archipelagos} when a 1-cell dilation connects them and \emph{islets} otherwise. Continents and islands are the \emph{landmasses}. Each feature inherits the dominant HDBSCAN~\cite{hdbscan} topic of its constituent documents.  Feature color in the rendering follows a tiered palette whose cutoffs match the classification cutoffs above: continents draw from a 10-color saturated palette, islands from an 18-color muted palette, and archipelagos and islets collapse to gray, with a darker shade reserved for those features the storyline visits (Figure ~\ref{fig:pipeline}c, Figure~\ref{fig:teaser}). Landmasses summarize regions of high document density on the projected sphere. A coastline traces where the estimated density crosses the land threshold, and the spatial arrangement of landmasses follows from document proximity in the \((\mathrm{lat},\mathrm{lon})\) projection. The Earth metaphor contributes familiar navigational affordances such as poles, a prime meridian, and rotation, while the coordinates underneath remain those of the embedding geometry. Labels summarize the documents within each feature and serve as a reading aid.

\emph{Cost.} KDE is \(O(G N)\) (\(O((G{+}N)\log N)\) with a BallTree), connected components \(O(G)\), and the geodesic-monotone MCP is Dijkstra at \(O(E \log V)\). The full projection-to-MCP pipeline runs in 0.32\,s on one CPU core.

\subsection{Labeling}\label{sec:naming}
Each feature receives a short label of 2--5 words drawn from its own constituent documents. We run a small local instruction-tuned language model (Qwen2.5-1.5B-Instruct, $\sim$1\,GB on CPU) in two passes per feature: a \emph{propose} pass over the feature's five most central documents, then a \emph{verify} pass that refines the label against those already assigned. The model sees only out-of-domain format examples (e.g., \emph{Apollo Landing}) to prevent corpus leakage, and labels failing a short surface validator fall back to TF-IDF over document titles. 

\section{Case study: Cuba--U.S. relations, 2016--2021}\label{sec:case}
Narrative extraction from open-domain news lacks canonical ground truth~\cite{keith2023survey}; thus the extracted path is validated by the analyst's domain reading.

Our reference corpus is 540 English-language news articles spanning Obama's 2016 Havana visit through the July 2021 protests and the subsequent Mexican aid flotilla, represented as 1536-$d$ OpenAI \texttt{text-embedding-3-small} vectors~\cite{openai_embed_3small}. We pin SRC = ``\emph{Obama on Cuba: Differences Remain but Change in Sight}'' (doc 15) and TGT = ``\emph{3 Mexican ships taking fuel, medical aid and food to Cuba}'' (doc 460); the great-circle angle between their embeddings is $50^\circ$. These endpoints mark the start and end of the period of interest and encode the analyst's question about it. In general, an analyst selects endpoints from domain knowledge of the two events that frame the narrative of interest, or through a retrieval step such as a query or a recommendation over the corpus. Because the projection is deterministic and inexpensive to recompute, an analyst can revise the endpoints and inspect the resulting narrative interactively. The projected sphere yields 27 landmasses (8 continents and 19 islands); the remaining small components form 20 archipelagos and 43 singleton islets. Labels emitted by the naming pipeline were manually reviewed for accuracy.

The geodesic-monotone MCP traces a 6-document chain whose events span four landmasses (including the endpoints) and two archipelagos (Figure~\ref{fig:teaser}): SRC (\emph{Obama on Cuba}) $\to$ \emph{US-Cuba Relations Policy Review} $\to$ \emph{Cuba Tourism and Trade} (archipelago) $\to$ \emph{US Sanctions Against Cuba} (archipelago) $\to$ \emph{Venezuela Sanctions Crisis} $\to$ \emph{Mexican Aid Flotilla} (TGT).  Every step advances latitude, and thematic deviation from the prime meridian is readable as each waypoint's azimuthal offset. The unconstrained MCP on the same pair uses 8 documents across 7 features, reaching \emph{US Sanctions Against Cuba} and then detouring backward through a lower-latitude feature before advancing to TGT. An orthogonal COVID-themed hemisphere---\emph{China Virus Outbreak}, \emph{Coronavirus Pandemic}, \emph{Cuba's Medical Revolution}---sits roughly $180^\circ$ off the narrative meridian, reflecting its near-independence from the diplomatic storyline. 

\section{Discussion}\label{sec:discussion}
\emph{Globe and flat rendering.} The pipeline renders the projected sphere as an Earth-like globe, but a flat Mollweide unfolding of the same \((\mathrm{lat}, \mathrm{lon})\) scatter is equally valid and may be more appropriate where 3D rendering is unavailable. The two renderings differ in the geometric respects described below.

\emph{Pole compression.} In the Mollweide projection the poles are single points---all longitudes at \(\mathrm{lat}=\pm\pi/2\) collapse to \((0,\pm\sqrt{2})\)---but documents \emph{near} the poles are compressed in \(x\) proportionally to \(\cos\theta(\mathrm{lat})\), which vanishes at \(\mathrm{lat}=\pm\pi/2\).  Documents with very different longitudes (thematically distant from the geodesic) therefore crowd together near the poles on the flat map. On the globe the full angular separation between high-latitude documents is preserved, making thematic distinctions near the narrative endpoints readable.

\emph{Geodesic preservation.} Only the gnomonic projection maps all great-circle geodesics to straight lines; every pseudo-cylindrical projection, including Mollweide, distorts off-meridian geodesics. The SLERP prime meridian is preserved in both views by construction, but any MCP step that departs from it (thematic deviation) is rendered faithfully on the globe and with angle-dependent distortion on Mollweide.

\emph{Rotation as the natural pan.} Panning a flat map requires edge-wrap artifacts or a new choice of center, breaking any narrative that straddles the antimeridian; on the globe, rotation is a continuous isometry (\(SO(3)\) acting on \(S^2\)) that reorients without re-parameterizing the projection. Both render the same projection, and the choice between them depends on the viewing setting.

\emph{Limitations.} We present Information Terra as an early-stage projection technique and design. The case study illustrates the method on a single corpus and one endpoint pair, and broader evaluation across corpora and endpoints remains open. The projection depends on the endpoints an analyst supplies. The advantages we describe for the globe rendering are geometric and await a user study against the flat Mollweide, though the extraction substrate inherits user validation from the narrative map lineage~\cite{shahaf_metromaps,narrativemaps_csci,keith2023iui, keith2026semantic}.

\section{Conclusion}\label{sec:conclusion}
Information Terra reuses the embedding sphere as a narrative-anchored substrate: endpoints become poles, the SLERP geodesic is the prime meridian, and the resulting coordinates drive both the coherence graph and the rendering. Potential future works include an immersive rendering of Information Terra, multi-narrative overlays (the globe analog of Narrative Maps), and scaling across corpus size and embedding model.

\section*{Supplemental Material}
The interactive demo and source code for Information Terra are available at \url{https://huggingface.co/spaces/briankeithn/InformationTerra}.

\acknowledgments{%
This work was supported by ANID FONDECYT de Iniciación 11250039. Generative AI was used in a limited capacity: Writefull (Overleaf) for grammar and linguistic clarity, and Claude Code for demo programming. All analytical and editorial decisions were made and reviewed by the authors, who take full responsibility for this work.
}

\bibliographystyle{abbrv-doi-hyperref-narrow}
\bibliography{references}

\end{document}